\documentclass[openacc]{rstransa}

\usepackage{amsmath}
\usepackage{amssymb}
\usepackage{graphicx}

\usepackage[numbers,sort&compress]{natbib}

\newcommand*{\doi}[1]{\href{http://dx.doi.org/#1}{doi: #1}}

\usepackage{hyperref}
\usepackage{enumitem}

\usepackage{tcolorbox}
\definecolor{abstract-color}{cmyk}{0.04, 0.04, 0.12, 0.08}

\begin{document}

\title{Characterisation of shock wave signatures at millimetre wavelengths from Bifrost simulations}

\author{
Henrik Eklund$^{1,2}$, Sven Wedemeyer$^{1,2}$, Ben Snow$^{3}$, David B. Jess$^{4,5}$, Shahin Jafarzadeh$^{1,2}$ , Samuel D.T. Grant$^{4}$, Mats Carlsson$^{1,2}$ and Miko\l{}aj Szydlarski$^{1,2}$}

\address{$^{1}$Rosseland Centre for Solar Physics, University of Oslo, Postboks 1029 Blindern, N-0315 Oslo, Norway\\
$^{2}$Institute of  Theoretical Astrophysics, University of Oslo, Postboks 1029 Blindern, N-0315 Oslo, Norway\\
$^{3}$Centre for Geophysical and Astrophysical Fluid Dynamics, University of Exeter, Exeter, UK\\
$^{4}$Astrophysics Research Centre, School of Mathematics and Physics, Queen's University Belfast, Belfast, BT7 1NN, U.K.\\
$^{5}$Department of Physics and Astronomy, California State University Northridge, Northridge, CA 91330, U.S.A.}

\subject{astrophysics, computational physics, computer modelling and simulation, solar system}

\keywords{shock waves, methods: numerical, Sun: chromosphere, Sun: photosphere, Sun: radio radiation}

\corres{Henrik Eklund\\
\email{henrik.eklund@astro.uio.no}}



\begin{fmtext}

\end{fmtext}


\maketitle

\begin{tcolorbox}[sharp corners, width=\textwidth,colback=abstract-color,colframe=abstract-color,boxsep=5pt,left=0pt,right=0pt,top=0pt,bottom=0pt]
Observations at millimetre wavelengths provide a valuable tool to study the small scale dynamics in the solar chromosphere. 
We evaluate the physical conditions of the atmosphere in the presence of a propagating shock wave and link that to the observable signatures in mm-wavelength radiation, providing valuable insights into the underlying physics of mm-wavelength observations.
A realistic numerical simulation from the 3D radiative Magnetohydrodynamic (MHD) code Bifrost is used to interpret changes in the atmosphere caused by shock wave propagation. 
High-cadence (1~s) time series of brightness temperature (T$_\text{b}$) maps are calculated with the Advanced Radiative Transfer (ART) code at the wavelengths $1.309$\,mm and $1.204$\,mm, which represents opposite sides of spectral band~$6$ of the Atacama Large Millimeter/submillimeter Array (ALMA).
An example of shock wave propagation is presented. 
The brightness temperatures show a strong shock wave signature with large variation in formation height between $\sim0.7$ to $1.4$\,Mm.
The results demonstrate that millimetre brightness temperatures efficiently track upwardly propagating shock waves in the middle chromosphere.
In addition, we show that the gradient of the brightness temperature between wavelengths within ALMA band~$6$ can potentially be utilised as a diagnostics tool in understanding the small-scale dynamics at the sampled layers.
\end{tcolorbox}


\section{Introduction}

The solar atmosphere is highly dynamic at small scales at chromospheric heights, also under quiet-Sun conditions with low magnetic-field strength \citep{1997ApJ...486L..63C}.
A major contribution to the small scale dynamics comes from the propagation of shock waves. Acoustic waves propagating upwards from the solar surface steepen into shock waves as a result of the decrease in gas density with height. The formation of shock waves and their propagation through the atmosphere have been studied by means of detailed one-dimensional numerical simulations, e.g., \citet{1970SoPh...12..403U, 1973ApJ...186.1083S, 1976A&A....47...65K, 1978A&A....70..487U, 1982ApJ...258..393L, 1991mcch.conf.....U, 1992ApJ...397L..59C, 1994chdy.conf...47C, 1997ApJ...481..500C, 2019A&A...626A..46S, 2020A&A...637A..97S, 2004A&A...419..747L, 2006A&A...456..713L, 1993A&A...273..671F}.

Three-dimensional (3D) simulations, for instance those by \cite{2004A&A...414.1121W},
exhibit a dynamic mesh-like pattern of hot filaments from shock waves surrounding cooler post-shock regions. Such 3D simulations are also employed by \cite{2007A&A...471..977W}
and \cite{2015A&A...575A..15L}
to explore the use of millimetre and submillimetre wavelengths as diagnostic tools for the chromosphere. 

The complex dynamics in the chromosphere have also been reported in numerous observational studies. In particular, small-scale (on the order of 1.5 Mm and smaller) and short-lived ($\approx100$~s or less) bright structures have been observed in the Ca~II~H and K spectral lines (the so-called H$_{2V}$/K$_{2V}$ grains), in agreement with the mesh-like patterns seen in the simulations \citep{1991SoPh..134...15R,1993A&A...274..584K,1996A&A...308..192H,2006A&A...459L...9W}, which are often interpreted as shock signatures \citep{2008A&A...479..213B}. While \cite{1982SoPh...80..227S} reported a correlation between these small-scale structures and a weak magnetic field of about 20~G, \citet{1993ApJ...414..345L} and \citet{2003ApJ...586.1409C} found no clear relationship between magnetic fields and the K$_{2V}$ grains. Other observational studies of small-scale shock signatures have also found that the magnetic field activity and orientation may play a major role in quiet Sun regions \citep{2007ASPC..368..127C,2009A&A...494..269V}, where shock waves propagate in both weak (or non-magnetised) and strong field-concentrated regions.
\citet{2008ApJ...680.1542H} showed that shock waves can produce excess temperatures of about 900~K in small magnetic concentrations in the chromosphere, which is responsible for the excess brightness observed in, e.g., small Ca II H magnetic bright points \citep{2013A&A...549A.116J}.

The radiation observed at millimetre wavelengths originates at chromospheric heights from free-free emission in local thermal equilibrium (LTE; see, e.g., \cite{2016SSRv..200....1W} and references therein). Following the Rayleigh-Jeans law \citep[i.e.,][]{2013tra..book.....W}, the measured intensities depend linearly on the local plasma temperature. In order to detect the small scale dynamics in observational mm-wavelength data, high spatial and temporal resolution is essential. In that regard, the Atacama Large millimetre/submillimetre Array (ALMA), which offers regular observations of the Sun since 2016, represents a major step forward in terms of resolution, and has already provided insights into the dynamics of mm-wavelength intensities, e.g., 
\cite{2017ApJ...841L...5S, 2018A&A...619L...6N, 2020arXiv200407591N, 2020A&A...635A..71W, 2020A&A...634A..86P,Jafarzadeh2020,Guevara-Gomez2020}.
Modelling has shown that propagating shock waves will cause variation in mm-wavelength observables
\cite{2004A&A...419..747L, 2006A&A...456..713L}.
However, these studies employed 1D models, therefore it is uncertain to what degree the intensity variations are affected by more realistic interactions of shock waves in a 3D environment. 

In this work, we use a 3D model that takes into account essential physical processes such as non-LTE and non-equilibrium hydrogen ionisation that have a large impact on the mm-wavelength radiation. With support from more realistic 3D simulations, it is possible to connect the mm-wavelength observables to the underlying physics and, thus, determine and characterise the observable signatures in mm-wavelengths of propagating shock waves.

The structure of the work is as follows. In Sect.\,\ref{sec:methods}, the setup of the simulations is explained and in Sect.\,\ref{sec:results} a representative example of a propagating shock wave with the surrounding physical conditions and the resulting signatures in brightness temperature are presented. In Sect.\,\ref{sec:disc}, we discuss the results and in Sect.\,\ref{sec:conclusions} conclusions are drawn and motivation for future work is given.


\section{Simulation setup}\label{sec:methods}

A three-dimensional numerical model of the solar atmosphere is created with the radiative magnetohydrodynamic (MHD) code Bifrost
\citep{2016A&A...585A...4C, 2011A&A...531A.154G}. 
The duration of the considered simulation sequence is approximately $1$\,hour with an output cadence of $1$\,s (matching the highest ALMA cadence in solar mode), so that rapid small-scale events on scales down to a few seconds can be efficiently studied. 
The simulation box has an extent of $24\times24\times17$\,Mm in horizontal ($x,y$) and vertical ($z$) directions, respectively.
The number of cells in $x$ and $y$ are both $504$, with a constant grid spacing of $48$\,km (corresponding to approx.  $0.066''$ at $1$\,AU). In the $z$-direction,  there are $496$ cells with grid spacing varying between $19$ -- $100$\,km, with a spacing of $20$\,km at chromospheric heights.
In the horizontal directions, the boundary conditions are periodic. 
The bottom boundary, which is located $2.5$\,Mm under the photosphere, allows flows (e.g. intergranular downdrafts) through, however, the average horizontal pressure is driven towards a constant value at a characteristic time of $100$\,s. This gives rise to acoustic wave reflection, mimicking the refraction of waves in the solar convection zone and giving rise to $p$-modes in the simulation. The upper boundary condition is based on characteristic equations and allows for the transmission of magneto-acoustic waves.

The simulation takes into account non-LTE and non-equilibrium hydrogen ionisation, as Hydrogen is the major contributor to the number of free electrons. Ionisation of other elements are under the assumption of LTE and, thus, given as function of internal energy and total mass density.
The simulation represents an 'enhanced network' region surrounded by quiet Sun \citep{2016A&A...585A...4C}.
The magnetic-field strength has an unsigned average value of $50$\,G ($5$\,mT) at photospheric heights with two opposite polarity regions of magnetic field approximately $8$\,Mm apart. 

The magnetic field was introduced into a relaxed convection simulation as two patches of opposite polarity at the lower boundary, with the rest of the simulation volume filled with a potential field extrapolation. This initial configuration is quickly modified by the convective flows that sweep the field into the intergranular lanes. The convective flows also cause foot point motions of the loops connecting the two polarities, leading to heating of the chromosphere and corona. The analysis of the simulation is restricted to later times (after 200s), when quasi-equilibrium has been established. For details of the setup and thermodynamic evolution of the simulation, see \cite{2016A&A...585A...4C}.

The same simulation with a cadence of $1$\,s has been used previously in the study by \cite{2020A&A...635A..71W}, to determine the diagnostic potential of solar ALMA observations. Other versions of the same simulation setup, but with lower cadence have been used in \cite{2015A&A...575A..15L, 2017A&A...601A..43L}, where the authors study chromospheric diagnostics at mm and sub-mm wavelengths with focus on the thermal structure and magnetic field.

The observable intensity at mm-wavelengths is obtained by solving the radiative transfer equation column by column. In LTE, the Rayleigh-Jeans approximation (valid for mm wavelengths) gives

\begin{equation}\label{eq:Tb}
T_\text{B}=\int_{-\infty}^{\infty}\chi_\nu T_\text{g} e^{-\tau_\nu} dz \ ,
\end{equation}

where $T_\text{B}$ is the brightness temperature, $\chi_\nu$ is the opacity, $T_\text{g}$ is the gas temperature and $\tau_\nu$ is the optical depth at height $z$ defined from
\begin{equation}\label{eq:tau}
\tau_\nu(z)=\int_z^{\infty}\chi_\nu dz \ .
\end{equation}

The integrand in equation (\ref{eq:Tb}) is the contribution function, describing which regions along the line-of-sight that contribute to the observed brightness temperature. We use the Advanced Radiative Transfer (ART) code (de la Cruz Rodriguez et al., in prep.) to solve the radiative transfer equation. The code assumes LTE but includes in detail the relevant sources of continuum opacity. For ALMA wavelengths, the opacity is dominated by free-free processes of hydrogen and H$^-$ (Loukitcheva et al 2004).

Two specific wavelengths are used in this study, located at the respective sides of the ALMA spectral band 6 in solar observing mode.
The ALMA receiver band 6 is sub-divided into four sub-bands: SB1 ($1.298$ -- $1.309$\,mm), SB2 ($1.287$ -- $1.298$\,mm), SB3 ($1.214$ -- $1.224$\,mm) and SB4 ($1.204$ -- $1.214$\,mm).
The wavelengths used here are $1.309$\,mm ($229.0$\,GHz) and $1.204$\,mm ($249.0$\,GHz), which are at the edges of SB1 and SB4, respectively. 
The brightness temperature, T$_b(\nu$), is calculated from the radiative intensities, I($\nu$), through the Rayleigh-Jeans law approximation.

\begin{figure}[!h]
\centering\includegraphics[width=3.5in]{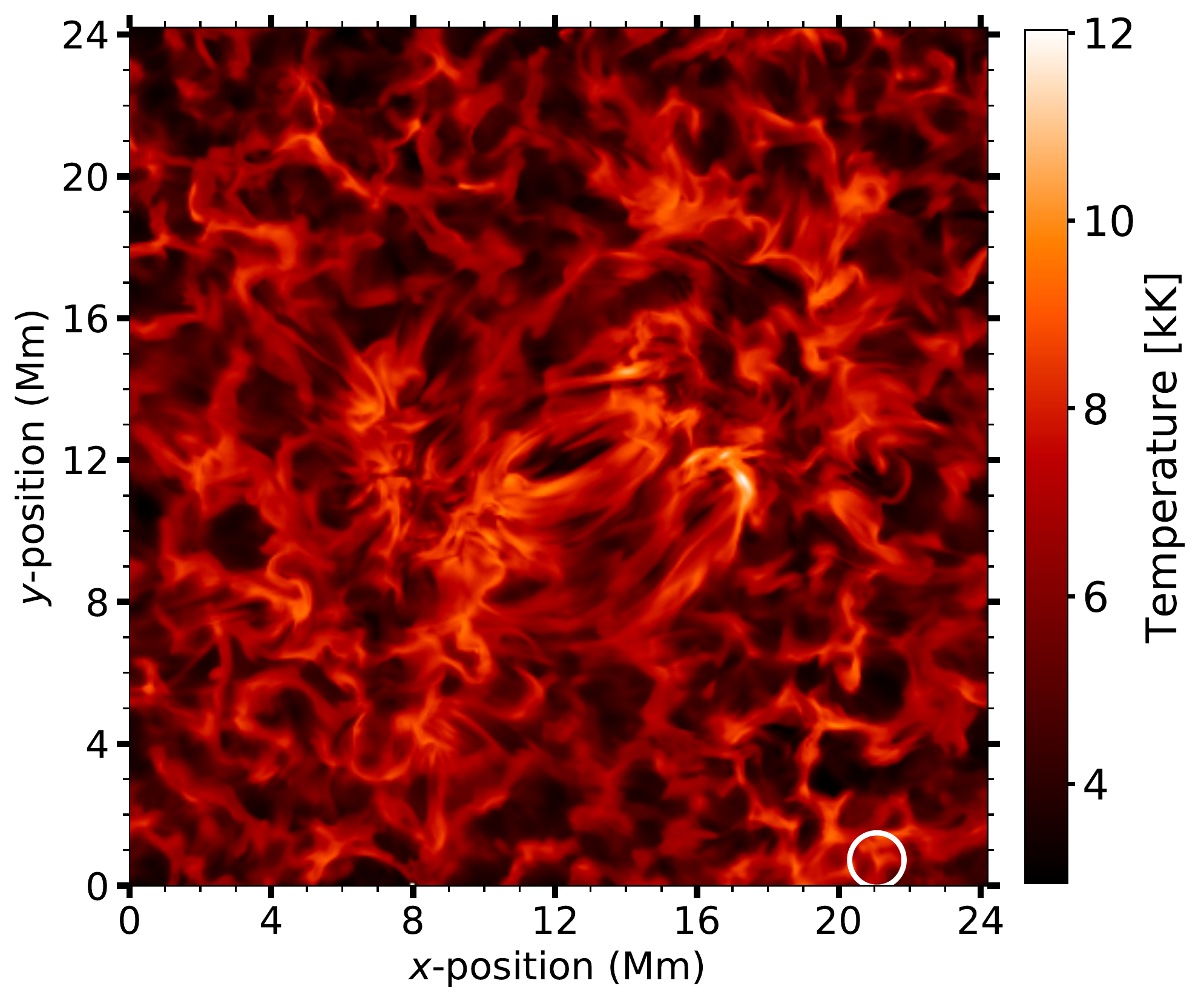}
\caption{Brightness temperature in SB1 ($1.309$\,mm; $229$\,GHz) at a time of $200$\,s after the beginning of the simulation. The white circle marks the location of the selected shock wave which is shown in more detail in Fig.\,\ref{fig_tg_xy_height}}.
\label{fig_FOV}
\end{figure}

Figure\,\ref{fig_FOV} shows the brightness temperature for SB1 at a time of $200$\,s from the start of the simulation. Over the whole $\sim1$\,h of simulation time, there are many signatures associated with shock waves. 
An example of a shock wave event, further described in detail in Sect.\,\ref{sec:results}, is located in the white circle marked in Fig.\,\ref{fig_FOV}, at $(\text{x},\text{y})=(21.07, 0.72)$. The event is visualised in more detail in Fig.\,\ref{fig_tg_xy_height}.
This example is representative of a typical shock wave in the simulation box, whose characteristics are presented from a qualitative point of view.

\section{Example of shock wave}\label{sec:results}

\begin{figure}[!h]
\centering\includegraphics[width=5.5in]{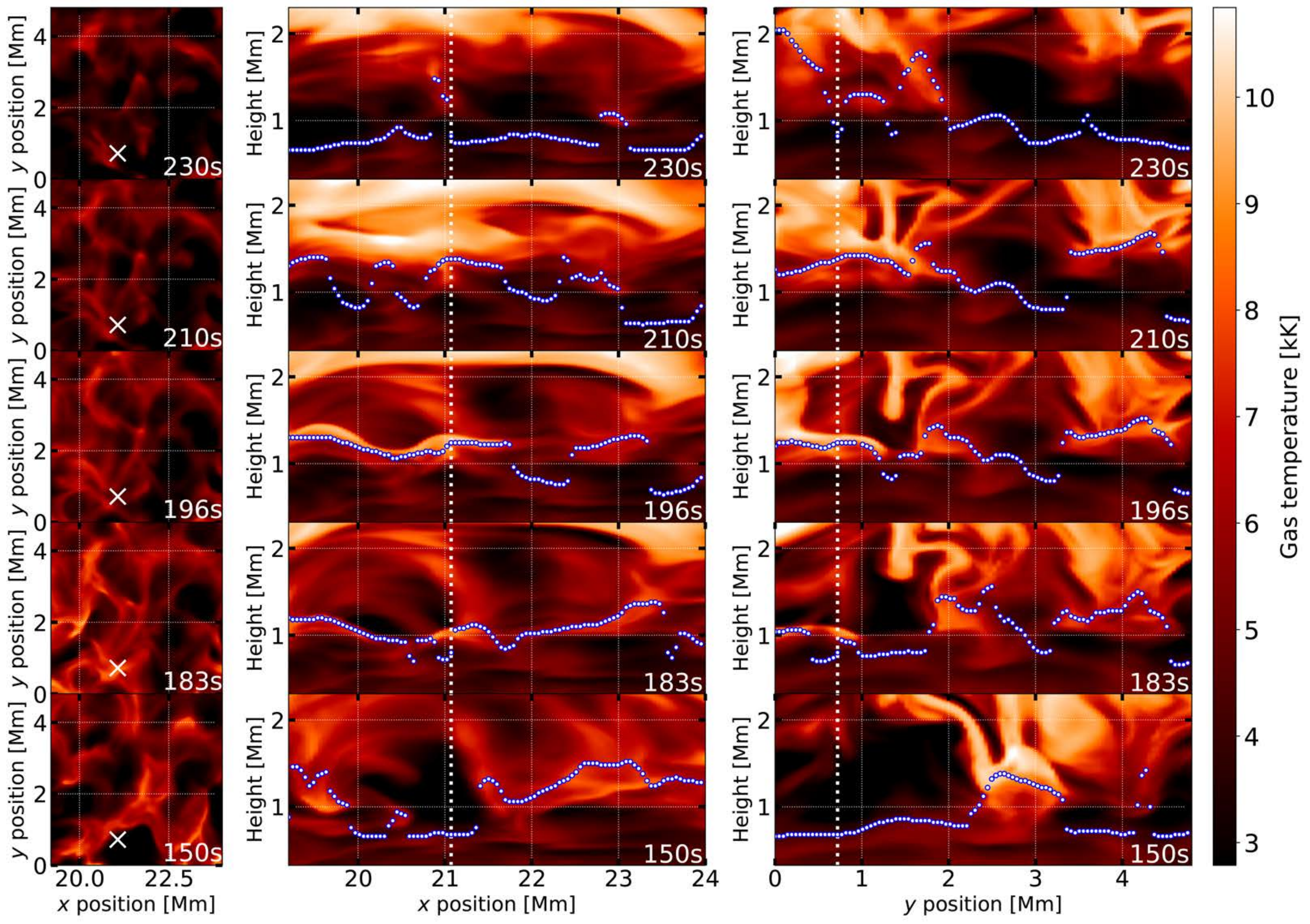}
\caption{Gas temperature surrounding the example of shock wave formation for five different time steps, from bottom to top: $t=150$, $183$, $196$, $210$ and $230$\,s. First column: Horizontal cuts at a fixed height of $1$\,Mm above the photosphere. The white cross marks the coordinates of the shock wave, $(x,y)=(21.1, 0.72)$\,Mm, which propagates largely in a vertical direction at these coordinates.
Second and third column: Vertical cuts through this position along the $x$-- and $y$--axis, respectively, showing heights between $\sim0.4$ -- $2.3$\,Mm. The white dotted lines marks the respective coordinate of the shock wave example. The blue dots show the height of optical depth $\tau=1.0$ for SB1.}
\label{fig_tg_xy_height}
\end{figure}

The regions surrounding the shock-wave event can be seen in the horizontal and vertical cuts of the gas temperature along the $z$-, $x$- and $y$-axis in Fig.\,\ref{fig_tg_xy_height}. The vertical cuts are given for five time steps spread out over the time span of the shock event. The bottom row shows the pre-shock phase at $t=150$\,s, dominated by cooler down-flowing gas. In the second row from the bottom, at $t=182$\,s, the shock has formed and reached a height of $\sim 1$\,Mm. In that moment, the brightness temperature is already at half its maximum value. 

The third row shows the peak phase of brightness temperature at $t=196$\,s, where the shock front reaches a height of approximately $1.25$\,Mm. The fourth row from the bottom shows the time $t=210$\,s when the brightness temperature is at maximum for this location. At this time the shock wave front has reached above $1.5$\,Mm. Finally, in the top row at $t=230$\,s, the shock wave front has reached well above $2$\,Mm and the cool post-shock medium is evident at formation heights of SB1 and SB4 around $\sim0.8$\,Mm.

By comparing the cuts between the time steps, it is possible to see the propagation of the shock wave to a height of $z\approx2.3$\,Mm. The shock wave propagation is predominantly vertical along the $z$-axis, except for a small tilt along the $y$-axis that can be seen in the rightmost column of Fig.\,\ref{fig_tg_xy_height}.

The ambient medium is highly dynamic, with a complex structure that leads to interactions between events. For instance, as the shock wave propagates through the chromosphere, background waves and structures affect the shock front.
The pre-shock medium, here seen around $t=150$\,s (Fig.\,\ref{fig_tg_xy_height}), is the resulting post-shock medium from preceding shock waves. Thus, the evolution of the shock wave front is dependent on how previous shock events influence the atmosphere. For this reason, it is necessary to use a realistic 3D atmospheric model in place of one-dimensional models in order to make predictions of the mm-wavelength signatures. The mm-wavelength radiation (marked in blue for $1.204$\,mm in Fig.\,\ref{fig_tg_xy_height}) is also dependent on the local atmospheric conditions, with a range of formation heights between $\sim0.6$ -- $2$\,Mm.


\subsection{Contribution function to brightness temperature}\label{sect:results_cf}

\begin{figure}[!h]
\centering\includegraphics[width=5.5in]{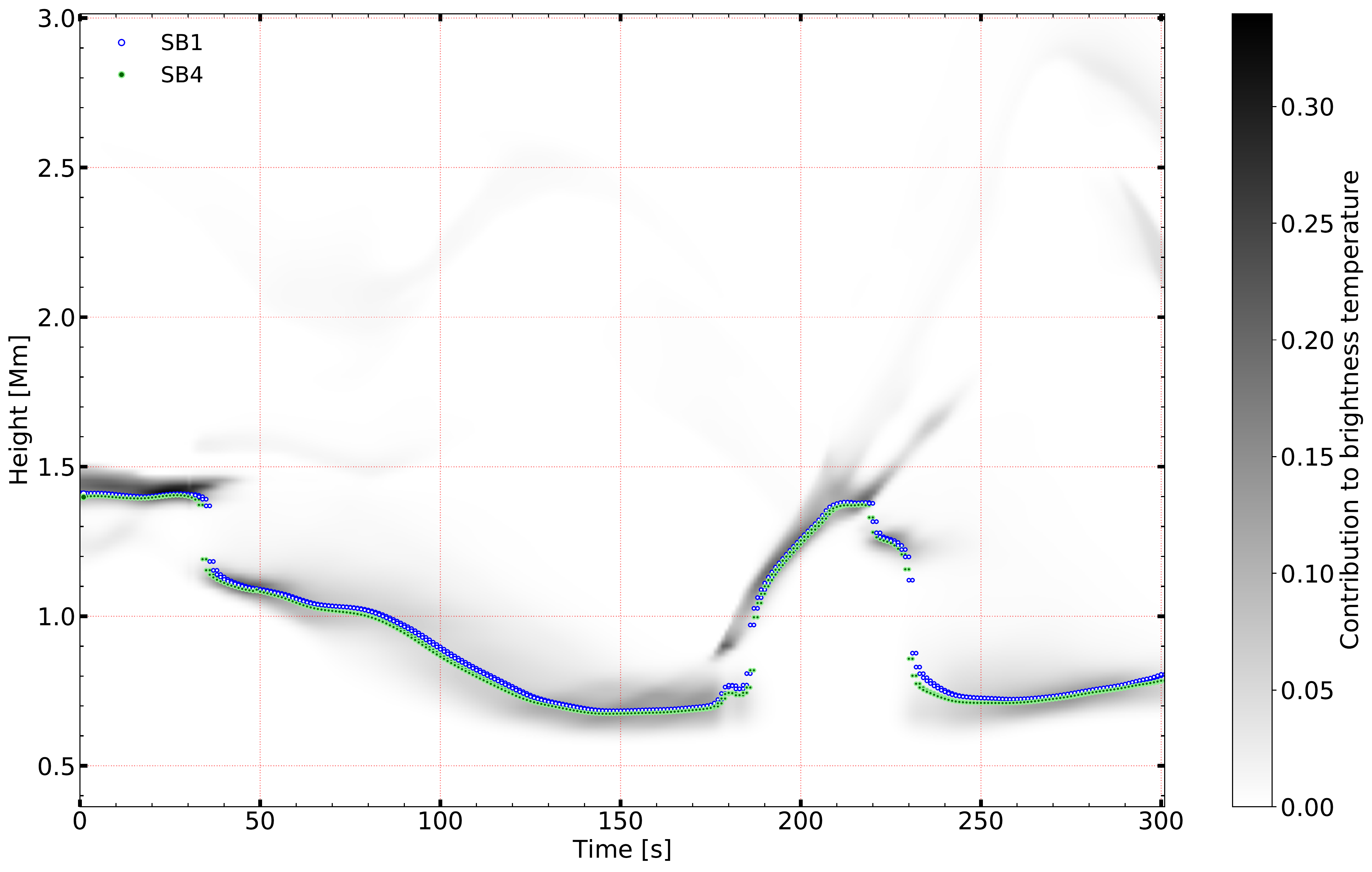}
\caption{Time evolution of the contribution function for SB1 at $1.309$\,mm. (The contribution function for SB4, at $1.204$\,mm, differs slightly but looks nearly visually identical at these scales). For each time step, the heights of $\tau=1.0$ for SB1 and SB4 are indicated by the blue and green markers, respectively. The integrated contribution function is normalised to $1.0$.}
\label{fig:cf}
\end{figure}

The time-dependent contribution function of the brightness temperature of SB1 ($1.309$\,mm) of the selected shock wave (see the location marked in Fig.~\ref{fig_FOV}) is given in Fig.\,\ref{fig:cf}. The corresponding contribution function for SB4 ($1.204$\,mm), is similar, however, there are small differences resulting in differing heights of optical depth $\tau=1.0$, marked by the blue and green dots in Fig.\,\ref{fig:cf}. A value for the optical depth of $\tau=1.0$ is often a good one-point approximation to the full contribution function.  As the shock wave front propagates upwards between the heights $z \approx 1.1$ -- $1.3$\,Mm (at $t\sim 190$ -- $210$\,s), the span of the contribution function is very narrow. Here, the formation height that $\tau=1.0$ corresponds to shows a strong correlation to, and effectively samples, the local gas temperature at a thin layer around the shock front. Immediately before $t=190$\,s, but mostly evident after $t=210$\,s, the brightness temperatures are sampled from several components at different heights. As the shock front propagates above $z\approx1.5$\,Mm, it is almost completely transparent in mm-wavelengths. There is a small fraction of the total contribution, no more than a few percent, that comes from the shock front at these heights, visible as a light grey streak in Fig\,\ref{fig:cf}. In the pre- and post-shock regimes, the contribution function spans a larger extent of heights.

\subsection{Gas and brightness temperature}\label{sect:results_temperatures}

\begin{figure}[!h]
\centering\includegraphics[width=5.5in]{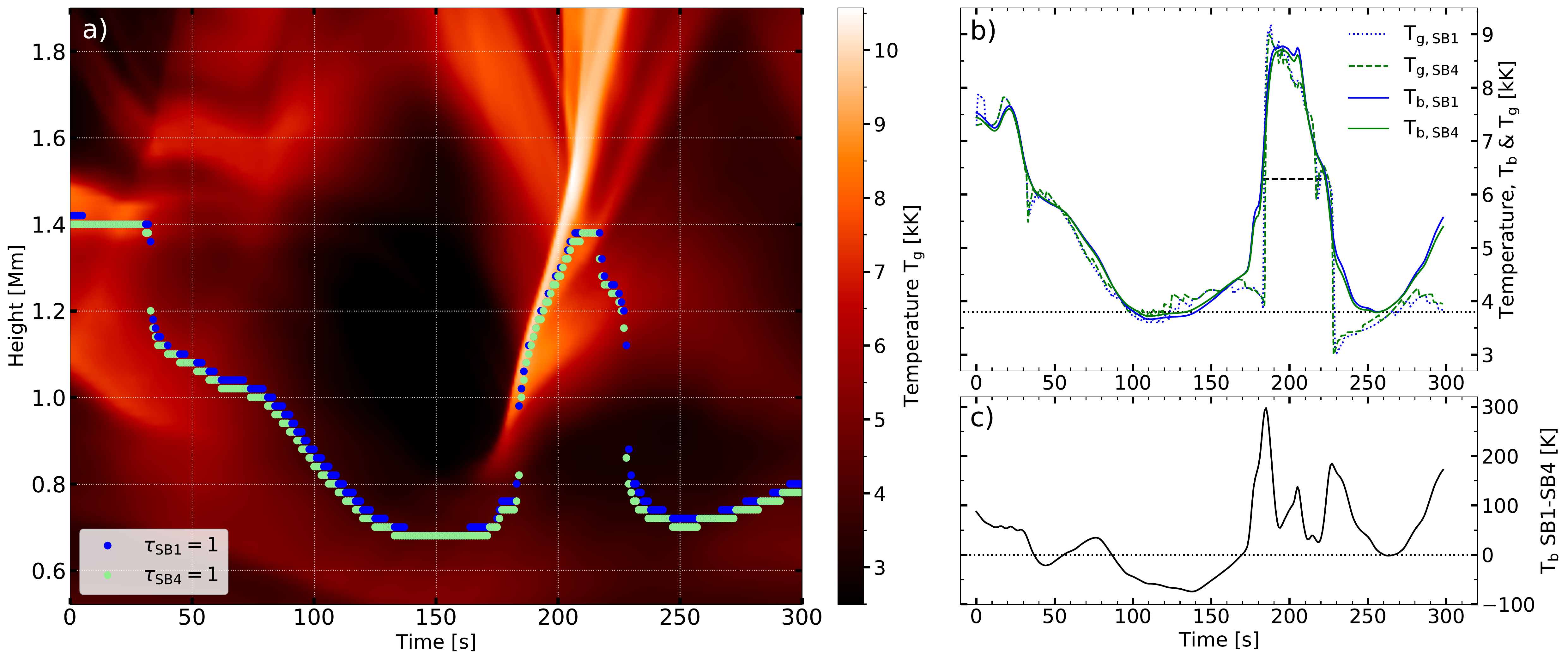}
\caption{Time-dependent temperatures in the selected shock wave example. \textbf{a)}~Evolution of gas temperature for one column at chromospheric heights between $\sim0.6$ and $1.8$\,Mm. The blue and green dots mark the formation heights (i.e., $\tau=1.0$) for the wavelengths $1.309$\,mm (SB1) and $1.204$\,mm (SB4), respectively.  
\textbf{b)}~Evolution of the brightness temperatures (T$_\text{b}$) at wavelengths of $1.309$\,mm (SB1, blue solid) and $1.204$\,mm (SB4, green solid) and of the gas temperatures (T$_\text{g}$) at heights corresponding to the optical depth of unity at the respective wavelengths (blue/green dotted). The horizontal dashed black line represents the Full Width Half Maximum (FWHM) lifetime of the brightness temperature shock wave signature (with respect of the base temperature, dotted horizontal line). \textbf{c)}~Evolution of the difference between the brightness temperatures of SB1 and SB4.}
\label{fig_tg_ev}
\end{figure}

The time evolution of the gas temperature during the propagation of the selected shock wave is shown in Fig.\,\ref{fig_tg_ev}a  for heights between $z\approx0.6$ -- $1.8$\,Mm. 
In addition, the heights where the optical depth, $\tau$, is unity at the wavelengths $1.309$\,mm (SB1) and $1.204$\,mm (SB4) are marked as a function of time. The wave steepens into a shock around t=$175$\,s, close to a height of $z=0.8$\,Mm and thereafter shows a very rapid increase in height. This height for shock wave steepening is consistent with other studies \citep[e.g.,][]{2004A&A...414.1121W}. The formation height varies largely with time as the shock wave propagates through the chromosphere. There are small differences in the formation height between SB1 and SB4, of up to $\sim40$\,km with a median value of $20$\,km, although they follow the same trend and keep the same order. As can be seen in Fig.~\ref{fig_tg_ev}a, the height where $\tau=1.0$ for both SB1 and SB4 increases from a pre-shock minimum of $z\approx0.68$\,Mm to a peak value of $\sim1.38$\,Mm during the course of $44$\,s with the propagation of the shock wave. 
The formation heights (i.e., $z(\tau=1)$) thus increase by $\Delta z \sim0.7$\,Mm from the low to the middle chromosphere during the upward propagation of the shock wave.

From this point on, the formation heights no longer follow the upward propagating shock front. Instead, after about $10$\,s at the peak height, the formation heights rapidly decrease to $z=0.72$\,Mm in just $27$\,s and thus map the post-shock phase. The brightness temperatures map the hot propagating shock wave front up to a certain height where it decouples as a result of the lower opacity at these heights for the mm-wavelengths of SB1 and SB4.

Fig.\,\ref{fig_tg_ev}b shows the corresponding evolution of the brightness temperatures of SB1 and SB4 as well as the gas temperature at the specific heights $z(\tau=1.0)$ for SB1 and SB4, respectively. The brightness temperature of SB1 shows a total increase of $\sim4970$\,K in $\Delta$t=$86$\,s, starting at a pre-shock local minimum of $3660$\,K at t=$110$\,s and rising to the peak value of $8780$\,K at t=$196$\,s. Thereafter, there is a rapid reduction to $3800$\,K at the local minimum at t=$256$ in the post-shock phase. The pre- and post-shock temperatures are thus comparable in this example. The local minimum with the highest temperature (here the post-shock minimum) is referred to as the `base temperature`. The time between the two local minima is $146$\,s. Estimating the lifetime of the observable brightness temperature signature as the Full Width Half Maximum (FWHM) of the peak results in  $t_\mathrm{FWHM}=41$\,s. The brightness temperature excess and lifetime shown here are in line with values derived from shock wave propagation in one-dimensional simulations (\cite{2004A&A...419..747L, 2006A&A...456..713L} and Eklund et al., in prep.). The resulting strong correlation between the gas temperature at $z(\tau=1.0)$ and the brightness temperature (Fig\,\ref{fig_tg_ev}b) is expected for mm-wavelength radiation (see i.e., \cite{2015A&A...575A..15L}). 

Figure~\ref{fig_tg_ev}c shows the time evolution of the difference between the brightness temperatures of SB1 and SB4. In the pre-shock phase around $85$ -- $170$\,s there is a difference of down to -$74$\,K, which corresponds to a magnitude of about 2\,\% of the total brightness temperature. Later during the shock phase, the difference between SB1 and SB4 increases to a total of 300\,K, corresponding to 4\,\% of the total T$_\text{b}$, and finally decreases to approximately zero in the post-shock phase.
The propagating shock wave and the pre-shock epoch display two different cases where the temperature gradient between the two sub-bands have a different sign. The brightness temperatures sampled at the propagation of the shock wave front show a positive gradient. That is, SB1 always forms in higher regions than SB4 and thus has a higher temperature. In contrast, during the pre-shock epoch, the gradient is negative and SB1 shows a lower temperature than SB4. The peaks in the time evolution of the brightness temperatures difference (Fig\,\ref{fig_tg_ev}c) centered around $\sim185$\,s, $200$\,s and $230$\,s originates from the signatures of three distinct wave components with differing propagation speeds. These are seen in the $t-z$ plot of Fig.\,\ref{fig_tg_ev}a as a hot, rapid component, followed by a cooler, slower component going upwards above $1.8$\,Mm and a third more diffuse component deflecting downwards around $1.2$\,Mm.


The rapid and large variations of the gas temperature at $z(\tau=1.0)$ (Fig.\,\ref{fig_tg_ev}b) are clearly connected to the large variations in formation height. The temporal profile of the brightness temperature (Fig\,\ref{fig_tg_ev}b) is integrated over a span of heights along the specific column and is therefore smoother than the gas temperature, which is sampled at a single height.


\subsection{Vertical velocity}\label{sect:results_velocities}
\begin{figure}[!h]
\centering\includegraphics[width=5.5in]{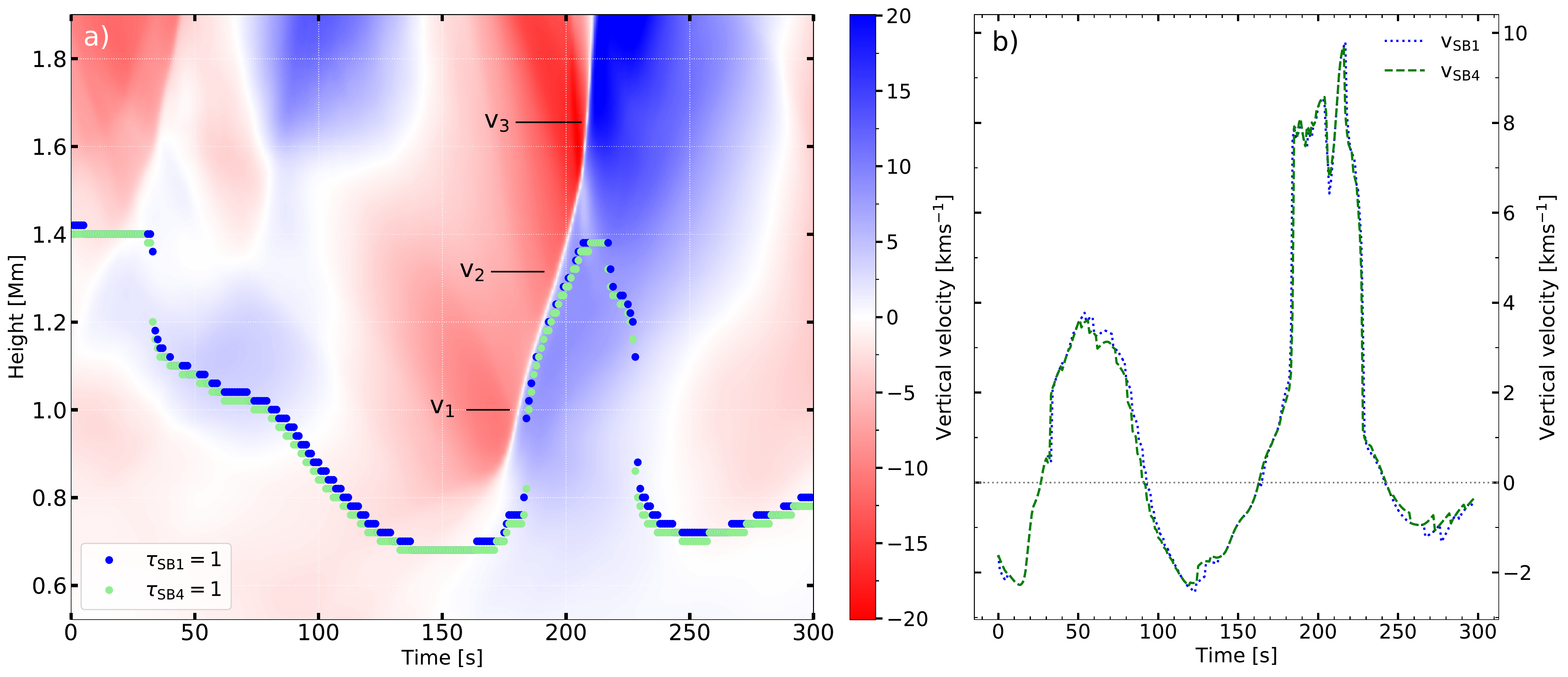}
\caption{Time-dependent vertical velocities for the selected shock wave example. \textbf{a)}~Evolution of vertical velocity for the chosen column (cf. Fig.\ref{fig_FOV}) at chromospheric heights between $\sim0.6$ and $1.8$\,Mm. Positive velocity (blue color) indicates outwards motion away from the photosphere. The markings $v_1$, $v_2$ and $v_3$ show the vertical propagation speed of the shock front at different heights.
The color scale is saturated on the positive side (from $23.7$ to $20.1$\,km\,s$^{-1}$). The blue and green dots mark the formation heights (i.e., $\tau=1.0$) for the wavelengths $1.309$\,mm (SB1) and $1.204$\,mm (SB4), respectively.
\textbf{b)}~Evolution of vertical velocity at heights corresponding to the optical depth unity at the wavelengths for  SB1 and SB4.}
\label{fig_vel_ev}
\end{figure}

The evolution of the vertical velocity ($v_z$) at the chosen position (cf. Fig. \ref{fig_FOV}) is shown in Fig.\,\ref{fig_vel_ev}a.
There is a bulk downflow of cooler gas (Fig.\,\ref{fig_tg_ev}a) in the pre-shock region, with velocities of up to $20$\,km\,s$^{-1}$. The shock front is met by this downfall and, therefore, experiences a resistance to its motion as it propagates upwards. In the height range from where the mm continuum radiation in  SB1 and SB4 originates, the vertical velocity only reaches a maximum velocity $v_z \sim 10$\,km\,s$^{-1}$, whereas in the upper chromosphere, there are velocities of up to $\sim20$\,km\,s$^{-1}$. In Fig.\,\ref{fig_vel_ev}a, the markings $v_1$, $v_2$, and $v_3$ point out the shock front at three different heights, $\sim 1.0, 1.3$ and $1.65$\,Mm. At these heights, the $t-z$~slope of the sharp transition of the vertical velocity (or the gas temperature in Fig.\,\ref{fig_tg_ev}a) indicates a speed of the vertical propagation of the shock wave of $\sim33$, $19$ and $83$\,km\,s$^{-1}$, respectively. 

The indications of differences in vertical velocity between heights of $z(\tau=1.0)$ for SB1 and SB4 are generally small (i.e., smaller than $1$\,km\,s$^{-1}$). This is a due to the height difference between $z(\tau=1.0)$ for SB1 and SB4 being of the same order as the vertical grid spacing of $20$\,km. 

\subsection{Electron density}\label{sect:results_electron_density}

\begin{figure}[!h]
\centering\includegraphics[width=5.5in]{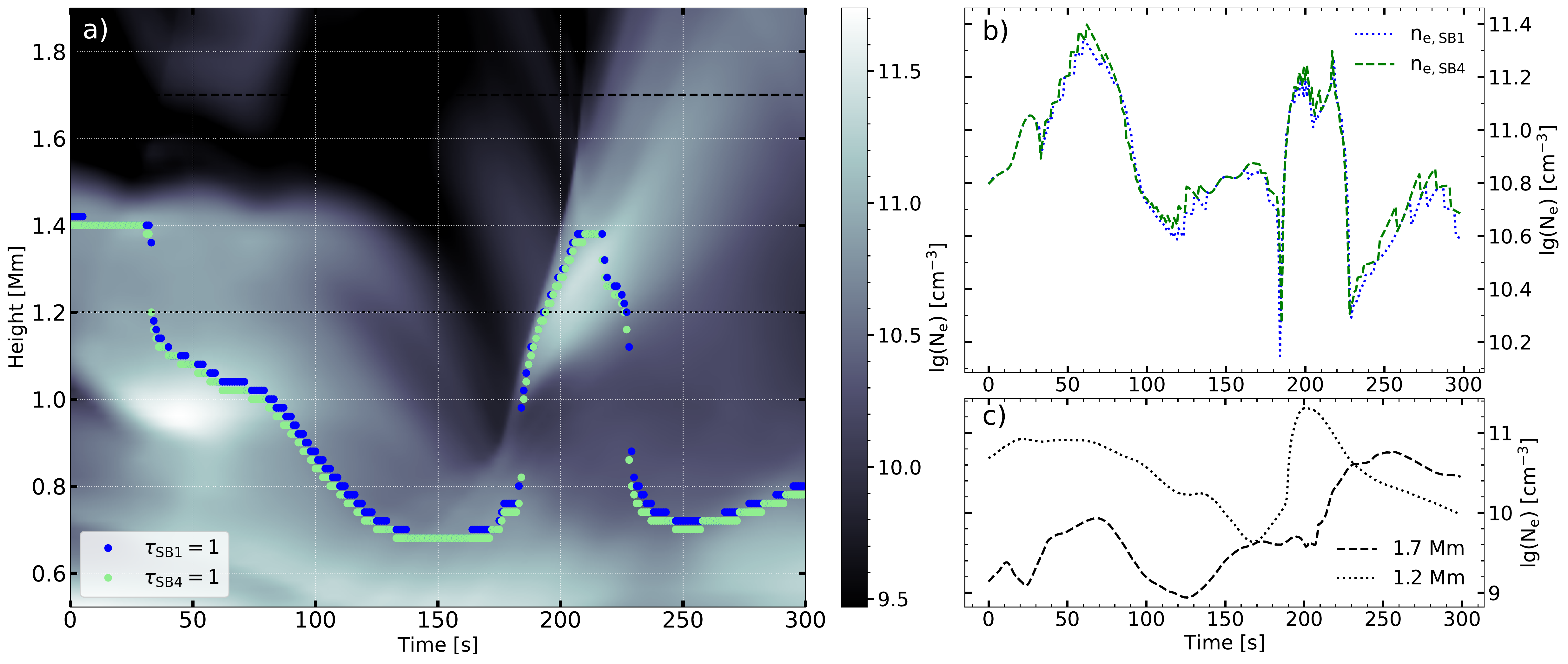}
\caption{Time-dependent electron density for the selected shock wave example. \textbf{a)}~Evolution of electron density for one column at chromospheric heights between $\sim0.6$ and $1.8$\,Mm. The blue and green dots mark the formation heights (i.e., $\tau=1.0$) for the wavelengths $1.309$\,mm (SB1) and $1.204$\,mm (SB4), respectively. The dotted and dashed horizontal lines marks the heights of $1.2$ and $1.7$\,Mm. \textbf{b)}~Evolution of electron density at heights corresponding to the optical depth of unity at the respective wavelengths of SB1 and SB4. \textbf{c)}~Evolution of electron density at fixed heights of $1.2$ and $1.7$\,Mm.}
\label{fig:ne_eveolution}
\end{figure}

The time evolution of the electron density for the column of the shock wave is given in Fig.\,\ref{fig:ne_eveolution}. 
During the pre-shock phase, the electron density is slowly decreasing in which the formation height of the brightness temperatures follows the same trend. There is a larger decrease followed by a rapid increase in the electron density around $\sim180$\,s as a result of the ionisation coming from the shock wave. In the post-shock regime, the electron density decreases slowly which combined with the declining gas temperature (Fig.\,\ref{fig_tg_ev}) at some point ($t~225$\,s) results in a sufficiently low opacity to initiate a sudden jump down in formation height of the mm-wave radiation.  
The correlation of the gas temperature and electron density in shocks \citep{2002ApJ...572..626C} is confirmed by the close relation of these quantities during the shock wave passage between $1.0$ and $1.3$\,Mm.

The local atmosphere shows an increased electron density for a significant time span after the shock wave propagates through. 
\cite{2002ApJ...572..626C} find through 1D models that the timescale of relaxation of the local atmosphere through hydrogen recombination after the shock wave has propagated through the chromosphere varies with height.  
They show a large span of timescales on the order of between $\sim10^2$ -- $10^5$\,s at chromospheric heights, with a peak value in the mid-chromosphere and rapidly decreasing towards both the photosphere and transition region. In Fig.\,\ref{fig:ne_eveolution}c, the time evolution of the electron density at two fixed heights, $1.2$ and $1.7$\,Mm are shown. The rate of decrease of electron density is slower at $1.7$\,Mm than at $1.2$\,Mm, for this shock wave. The relaxation times are difficult to measure due to the dynamic atmosphere with preceding, as well as succeeding, wave trains at the same position, ensuring that the electron density never reaches a steady state. Estimating the relaxation times by simply extrapolating with the same trend as for the last $30$\,s between $270$\,s and $300$\,s, results in $\sim200$\,s and $\sim430$\,s to reach values of previous minima at the heights $1.2$ and $1.7$\,Mm, respectively. 

The difference of the electron density between SB1 and SB4 is on the order of $\log(N_e)= 0.1$\,cm$^{-3}$. The vertical grid spacing is in the same order as the differences in formation height $z(\tau=1)$ of SB1 and SB4. Therefore, as with the velocities in Sect.\,\ref{sect:results_velocities}, it is difficult to make use of the differences between electron density of SB1 and SB4 that are sampled at $z(\tau=1)$.

The opacity of the mm-wavelengths is dominated by free-free processes (e.g., \cite{1985ARA&A..23..169D} and \cite{2004A&A...419..747L}) and is reduced by an increase in temperature coupled with a decrease in electron density.
This behaviour is seen as the shock wave propagates upwards, resulting in a decrease of the opacity. 
At some point the opacity is reduced to the level where the brightness temperature de-couples from the shock wave front, leading to a rapid decrease in the formation height $z(\tau=1)$.


\section{Discussion}\label{sec:disc}

The shock wave example illustrated in this work was found to be representative of a typical shock wave found in the simulation with respect to variations of gas temperature, vertical velocity and electron density. As a result of the complex 3D structure of the atmosphere, the propagation of the shock waves can be intricate, with differing speeds in different directions, alongside changes in the propagation angle of the wave-front. The shock wave example presented in this study exhibits a predominantly vertical propagation, which is to be expected for the region under consideration, with magnetic fields of minimal inclination. The temporal profile of a shock wave event will show deformations depending on the specific propagation properties at different locations (i.e., different columns). Thus, the vertical propagation of the shock front under consideration ensures a simple inference of the brightness temperature, absent of any large 3D components.

The formation height of mm-waves in the chromosphere is a function of wavelength, with height increasing with wavelength. Accordingly, SB1 forms slightly higher up than SB4. This is seen throughout the entire example of shock wave propagation in the chromosphere (Fig.\,\ref{fig_tg_ev}a), despite the presence of small-scale dynamics in the chromosphere.
The difference of $0.105$\,mm between SB1 and SB4 is shown to be enough to map different layers with brightness temperature differences up to $\sim300$\,K (Fig\,\ref{fig_tg_ev}c). There are however notable variations in the differential between SB1 and SB4, sometimes to such a high degree that a reversal of the sub-bands are evident. The strongest reversal is seen in connection to the down-falling cold gas seen in the pre-shock region ($\sim140$\,s) of the illustrated example.

With larger wavelength separation between the sampled sub-bands, larger T$_\text{b}$ differences would be observed as the sampled layers would lie further apart. For that reason, it would be of interest in regards of the temporal domain to track a propagating shock wave from one layer to the other. Observations with ALMA with increased separation between the sub-bands would be favourable. Solar ALMA observations are currently offered in several spectral bands between $\sim0.8$ -- $3.3$\,mm. ALMA band\,$3$ ($2.8$ -- $3.3$\,mm) offers the largest default separation between the outermost sub-bands of $0.42$\,mm, which is $\sim4.5$ times more than that of band\,$6$ in this study. However, further consideration needs to be made, such as the change in formation height and, thus, potential change of shock wave propagation speed and contrast of the dynamic signatures.

\subsection{Contribution function spread over geometric height}

Figure~\ref{fig:cf} reveals the important relationship between the developing shock and the associated $\tau=1.0$ region. Importantly, it can be seen that during the initial formation of the shock ($\sim180$~s in Fig.~\ref{fig:cf}), the plasma is both dense and bright, resulting in the observed signatures at the $\tau=1.0$ location being dominated by the shocked plasma. Initially, this relationship continues to hold as the shock develops and propagates into higher layers of the chromosphere. However, at $t\sim220$~s in Fig.~\ref{fig:cf}, it can be seen that the contribution function defining the $\tau=1.0$ surface begins to decouple from the upwardly propagating shock. At this point, the contribution function will be comprised of both shocked plasma expanding upwards into more diffuse and optically thin layers of the atmosphere, in addition to cooling plasma beginning to accelerate back towards the solar surface, which is visible in Figures.~\ref{fig_tg_ev}a and \ref{fig_vel_ev}a. This results in the contribution function being spread over multiple components spanning a vast assortment of geometric heights. At this point, the signatures extracted at the $\tau=1.0$ height no longer strictly correspond to the propagating shock, which has important implications for observational studies of such phenomena. For example, recent work surrounding shocks manifesting in sunspot umbrae \cite{2013A&A...556A.115D, 2020ApJ...892...49H} have described the challenges faced when interpreting the spectropolarimetric fluctuations in Stokes~$I$/$Q$/$U$/$V$ spectra over the lifetime of a shock. Hence, the opacity response of propagating shocks affects a wide range of observable signals, spanning brightness temperatures in the radio regime through polarimetric intensities across the visible and infrared spectrum. In particular, recent work by \cite{2020ApJ...892...49H} interpreted reversals in the Stokes~$Q$/$U$ spectra as evidence for the presence of intermediate shocks, but this interpretation relied upon the observed signals being closely coupled to the developing shock front. As such, future investigations of challenging shock signatures (e.g. \cite{2018ApJ...860...28H, 2018A&A...619A..63J, 2018NatPh..14..480G}) need to carefully consider the potential effects of the contribution function decoupling from the shock as it propagates into less opaque regions of the solar atmosphere.

\subsection{Observations compared to numerical simulations}

Large advances have been made in interferometric observations of the Sun in mm-wavelengths, with ALMA. There are, however, many challenges that come with solar ALMA observations, for example, but not limited to, image reconstruction of interferometric data, limited spatial resolution, absolute temperature measurements, atmospheric noise, etc., that are out of the scope of this work.

The sophisticated 3D simulations give realistic predictions of how the shock wave signatures would look in brightness temperature as observed at mm-wavelengths. This work considers the brightness temperatures calculated at a horizontal resolution element of $48$\,km ($\sim 0.066$ arcsec). Performing actual observations at these wavelengths comes with instrumental resolution limitations that need to be considered. Though further complications arise from the limited spatial resolution of observations, the results of this work point towards the optimistic capability of ALMA with highly resolved data. As a result of limited spatial resolution, the magnitude of the excess T$_\text{b}$ of the dynamic profiles will appear less strong due to the pixel filling factor comprising of both the shocked plasma and cooler, quiescent plasma. 

The estimated difference in formation height between the wavelengths $1.309$ (SB1) and $1.204$\,mm (SB4) of up to $40$\,km with a median value of $20$\,km is on the limit of the vertical grid spacing of the simulation. The vertical resolution of $20$\,km in the chromosphere puts a limit on the differences of the small scale dynamics that can be handled. To study the differences between the sub-bands in more detail, a higher resolved numerical model would be necessary.

Observations also come with a certain amount of noise. The signal-to-noise ratio needs to be high enough to accurately deal with the magnitude of T$_b$ variations of interest. 
A few studies of ALMA data have been made where T$_b$ variations of small scale structures have been reported \citep{2018A&A...619L...6N, 2020A&A...635A..71W, 2020arXiv200407591N}.
In these studies, the brightness temperatures of the full spectal bands (all four sub-bands combined) were used. Integration over larger spectral or temporal spans can be done in order to increase the signal-to-noise ratio. To accurately map the T$_b$ differences between two sub-bands introduces a larger uncertainty. There are studies where the sub-bands are successfully used separately \citep{2019ApJ...875..163R, 2019A&A...622A.150J}, which in this case acts as a proof of concept.
Detection of brightness temperature variations as small as $70$\,K has been reported by \cite{2020arXiv200407591N}, where they use ALMA observations at $\sim3$\,mm (band 3) with integration over the full band with a cadence of $2$\,s. The spatial resolution element of their band\,$3$ data ($2.5''\times 4.5''$), is larger than what currently can be achieved with band\,$6$ data around $1.204$--$1.309$\,mm. This is a direct result of the Fourier sampling (fringe spacing) of the interferometric data scales with wavelength \citep[i.e.,][]{2013tra..book.....W}. With regards to the ability to spatially resolve small scale events, the ability to measure precise brightness temperatures should therefore be even more precise in band\,$6$ than in band\,$3$.
However, the integration over the full band comes with the inherent loss of sampling different layers as a function of wavelength. There should be an optimal combination of improving upon the signal-to-noise ratio whilst sampling different layers within a spectral band, so that differences on the order of one hundred Kelvin can be detected, as indicated by the $T_\text{b}$ difference between the sub-bands in the simulations (Fig.\,\ref{fig_tg_ev}c). 

Estimating the observable signatures of shock waves in mm-wavelengths with current and potential future modes offered for solar-ALMA observations, including the effect of limited spatial resolution of different spectral bands and the sampling of different physics between the sub-bands will be investigated in a forthcoming paper.


\section{Conclusion}\label{sec:conclusions}

We use realistic numerical 3D MHD simulations from the Bifrost code, including non-LTE, non-equilibrium hydrogen ionisation, of the solar atmosphere to study small scale dynamics connected to propagating shock waves and how these are perceived in mm-wavelength radiation. 
An example of a shock wave with nearly vertical propagation and without much interference from neighbouring dynamical features is illustrated. 
The shock wave propagating upwards in the chromosphere at vertical velocities between $\sim19-83$\,km\,s$^{-1}$, and has an assosiated increase in the local gas temperature of the order of several thousand degrees.
We conclude that the brightness temperature at mm-wavelengths corresponding to ALMA band 6 ($1.204$ -- $1.309$\,mm) probes these gas temperatures accurately under the highly dynamical conditions arising from propagating shock waves. 
The gas temperature at a single height $z(\tau=1.0)$ is quite close to the brightness temperature, which demonstrates the close relationship and the diagnostic potential for determining actual gas temperatures from mm-wavelength observations. The FWHM lifetime of the T$_\text{b}$ shock wave signature is $41$\,s.

The formation height of the radiation at a certain wavelength is not fixed. The formation height of wavelengths $1.204$ -- $1.309$\,mm varies on the order of $\sim0.7$\,Mm, from $\sim0.7$ to $\sim1.4$\,Mm in less than a minute, in the course of the shock wave propagating through the chromosphere.
The brightness temperatures at wavelengths corresponding to ALMA band~$6$ at $1.204$ -- $1.309$\,mm efficiently maps the shock front while it is propagating from approximately $1.0$\,Mm up to $1.4$\,Mm, where the brightness temperatures start to decouple and instead starts to map the post-shock region. The shock wave front continues propagating upwards, unseen by radiation at $1.204$ -- $1.309$\,mm. 
In the pre and post-shock regimes, the radiative contribution function at these wavelengths is more diffuse and spread out over larger span of heights. At some instances, the brightness temperature (at one frequency) has contributions from distinct layers at different heights. This is the scenario right before and after the strong coupling of the brightness temperature with the shock front.

There is a wavelength dependency of the optical depth which has been explored for wavelengths lying in the furthest apart sub-bands of ALMA spectral band 6. 
The simulations indicate that the difference in formation height between wavelengths SB1 and SB4 is up to approximately $40$\,km with a median difference of $20$\,km. The order of the formation heights with SB1 forming higher up than SB4, is however constant. Because of the correlation between the formation height and the wavelength of radiation, the gradient of brightness temperatures within the spectral band corresponds to a gradient in plasma temperature between the respective formation heights.
The brightness temperatures of SB1 and SB4 show differences from about $-70$\,K up to $\sim300$\,K in the shock wave example.
The difference between the sub-bands comes from the local temperature gradient between the mapped layers at the formation heights of the sub-bands. As the brightness temperature is coupled to the shock wave front, SB1 ($1.204$\,mm) has a higher temperature than SB4 ($1.309$\,mm) and there is a positive gradient with increasing temperature with height. In the pre- or post-shock regimes dominated by sampling of cold down flowing gas, the temperature gradient tends to be negative with SB1 colder than SB4.

The presented simulation results demonstrate that brightness temperatures of wavelengths corresponding to ALMA spectral band 6 ($1.204$ -- $1.309$\,mm) can be used for tracking shock waves from the middle chromosphere and that the gradient of the brightness temperature within the spectral band in principle can be utilised as a diagnostics tool for probing the small-scale structure of the chromosphere.

\enlargethispage{20pt}


\dataccess{\\
The Bifrost simulation with $10$\,s cadence is publicly available at: http://sdc.uio.no/search/simulations}

\aucontribute{MS and MC performed the MHD simulations and radiative transfer computations. HE performed scientiﬁc analysis, with assistance from SW, BS, DBJ, SJ, SDTG, and MC. HE drafted the manuscript. All authors read and approved the manuscript.}

\competing{The authors declare that they have no competing interests.}

\funding{This work is supported by the SolarALMA project, which has received funding from the European Research Council (ERC) under the European Union's Horizon 2020 research and innovation programme (grant agreement No. 682462) and by the Research Council of Norway through its Centres of Excellence scheme, project number 262622 (Rosseland Centre for Solar Physics). 
The development of the Advanced Radiative Transfer (ART) code was supported by the PRACE Preparatory Access Type D program (proposal 2010PA3776). 
Grants of computing time from the Norwegian Programme for Supercomputing are acknowledged. BS is supported by STFC research grant ST/R000891/1. DBJ and SDTG are supported by an Invest NI and Randox Laboratories Ltd. Research \& Development Grant (059RDEN-1).}

\ack{DBJ and SDTG are grateful to Invest NI and Randox Laboratories Ltd. for the award of a Research \& Development Grant (059RDEN-1).
We also acknowledge collaboration with the Solar Simulations for the Atacama Large Millimeter Observatory Network (SSALMON, \href{http://www.ssalmon.uio.no}{http://www.ssalmon.uio.no}).
 The support by M. Krotkiewski  from USIT, University of Oslo, Norway, for the technical development of ART is gratefully acknowledged.
BS, DBJ, SJ, and SDTG wish to acknowledge scientific discussions with the Waves in the Lower Solar Atmosphere (WaLSA; \href{https://www.WaLSA.team}{www.WaLSA.team}) team, which is supported by the Research Council of Norway (project no. 262622) and the Royal Society (award no. Hooke18b/SCTM).}



\bibliographystyle{rstasj}

\bibliography{ms1}

\end{document}